\documentclass[letterpaper, 10 pt, conference]{ieeeconf}  % Comment this line out if you need a4paper

\IEEEoverridecommandlockouts                              % This command is only needed if 
\usepackage{xcolor}
\usepackage{latexsym,amsmath,amssymb,amsfonts,mathrsfs,mathtools}

\usepackage{epsfig} % for postscript graphics files

\title{\LARGE \bf
Dynamical model-based experiment design for drug repositioning*
}

\author{Atte Aalto$^{1}$, La Mi$^{1}$, Diego A. Blanco-Mora$^{1}$, and Jorge Gon\c{c}alves$^{1,2}$ % <-this % stops a space
\thanks{*LM is supported by the Luxembourg National Research Fund (FNR) project INTER/JPND/20/14609071/Dynasti. DAB is supported by the FNR project INTER/DFG/21/15020234/BIML-19.}% <-this % stops a space
\thanks{$^{1}$All authors are with Luxembourg Centre for Systems Biomedicine, University of Luxembourg, Luxembourg}%
\thanks{$^{2}$ Department of Plant Sciences, University of Cambridge, UK}%
\thanks{{\small Emails:} {\tt\small \{atte.aalto, la.mi, diego.blanco-mora, jorge.goncalves\}@uni.lu}}
}

\begin{document}

\maketitle
\thispagestyle{empty}
\pagestyle{empty}

%%%%%%%%%%%%%%%%%%%%%%%%%%%%%%%%%%%%%%%%%%%%%%%%%%%%%%%%%%%%%%%%%%%%%%%%%%%%%%%%
\begin{abstract}

Computational methods in drug repositioning can help to conserve resources. In particular, methods based on biological networks are showing promise. Considering only the network topology and knowledge on drug target genes is not sufficient for quantitative predictions or predictions involving drug combinations. We propose an iterative procedure alternating between system identification and drug response experiments. Data from experiments are used to improve the model and drug effect knowledge, which is then used to select drugs for the next experiments. Using simulated data, we show that the procedure can identify nearly optimal drug combinations.

\end{abstract}

%%%%%%%%%%%%%%%%%%%%%%%%%%%%%%%%%%%%%%%%%%%%%%%%%%%%%%%%%%%%%%%%%%%%%%%%%%%%%%%%
\section{INTRODUCTION}

Developing a new drug is a complex task that requires abundant resources. Repurposing already approved drugs can alleviate some of the challenges, but testing hundreds of drugs experimentally is still prohibitively expensive. Computational approaches, in particular, biological network-based methods, have emerged as a promising strategy to screen through several drugs \emph{in silico} and to find good drug candidates \cite{Alaimo,LoftiShahreza}.

A common problem formulation in network-based drug repositioning is to find target genes corresponding to the set of differentially expressed genes between a healthy and disease states \cite{network1,network2,network3}. The obtained list of target genes can then be aligned with drug target genes to identify potential drugs. These methods are based on graph theoretic concepts. A method based on control-theory concepts was introduced in \cite{decost}. They construct a discrete-time linear system based on a protein-protein interaction network database, with $A$-matrix entries set to $\pm 1$ depending on the type of regulation. This system is used to solve an optimal control problem, whose solution is compared with a drug database.

In this article, we follow the same paradigm of using the network to connect drug target genes with the differential expression pattern between healthy and disease states. However, methods based solely on the graph topology are limited to qualitative single-drug predictions. Two challenges stand in the way of quantitative predictions. Firstly, the system dynamics must be known in more detail than just the underlying regulatory network topology. Secondly, in drug databases, typically only the drug target genes are reported, but not the magnitudes of the effects. To overcome these challenges, we propose an iterative scheme that alternates between drug response experiments, model improvements, and drug effect predictions, that are, in turn, used to select drugs to be tested in the next iteration.

The goal is to find a drug combination that steers a disease system's steady state towards a pre-defined state corresponding to a healthy state. However, the system itself is not known, except for the topology of the underlying gene regulatory network, which can be obtained from a network database, such as OmniPath \cite{omni}, or inferred from data as a preliminary step \cite{bingo,riva}. Moreover, the drug effects are only partially known \emph{a priori}. The iterative process begins with a high-frequency experiment with densely sampled time points using a drug that is known to elicit a strong response. This initial experiment is used for gathering data for initial model fitting and to investigate the time scale of the drug response. This knowledge is used to select measurement times for subsequent lower frequency drug experiments. 

Based on the inferred model and knowledge of drug target genes in a drug library, such as DrugBank \cite{drugbank}, the effects of drugs on the model's steady state are estimated. A scoring scheme is used to rank drugs based on their importance in different drug combinations. Top-scoring drugs are selected for low-frequency experimental testing. Data from these tests are used to improve the model, and for quantitative estimation of the drug effects. This results in an iterative process between computational modelling and experimental testing. Availability of experimental resources determines how many iterations can be carried out, and how many drugs can be tested on one iteration.

Although in the present paper we only test the procedure \emph{in silico}, the assumed prior knowledge and lab resource constraints closely emulate real experimental conditions involving midbrain organoids in disease state, in anticipation of processing real data.

{\bf Notation:}  $\|x\|_0$ denotes the number of nonzero elements in $x$. We use superscripts in parentheses $X^{(k)}$, to denote variables (and functions) on the iteration $k$.

   \begin{figure*}[ht!]
      \centering
      \includegraphics[width=17cm]{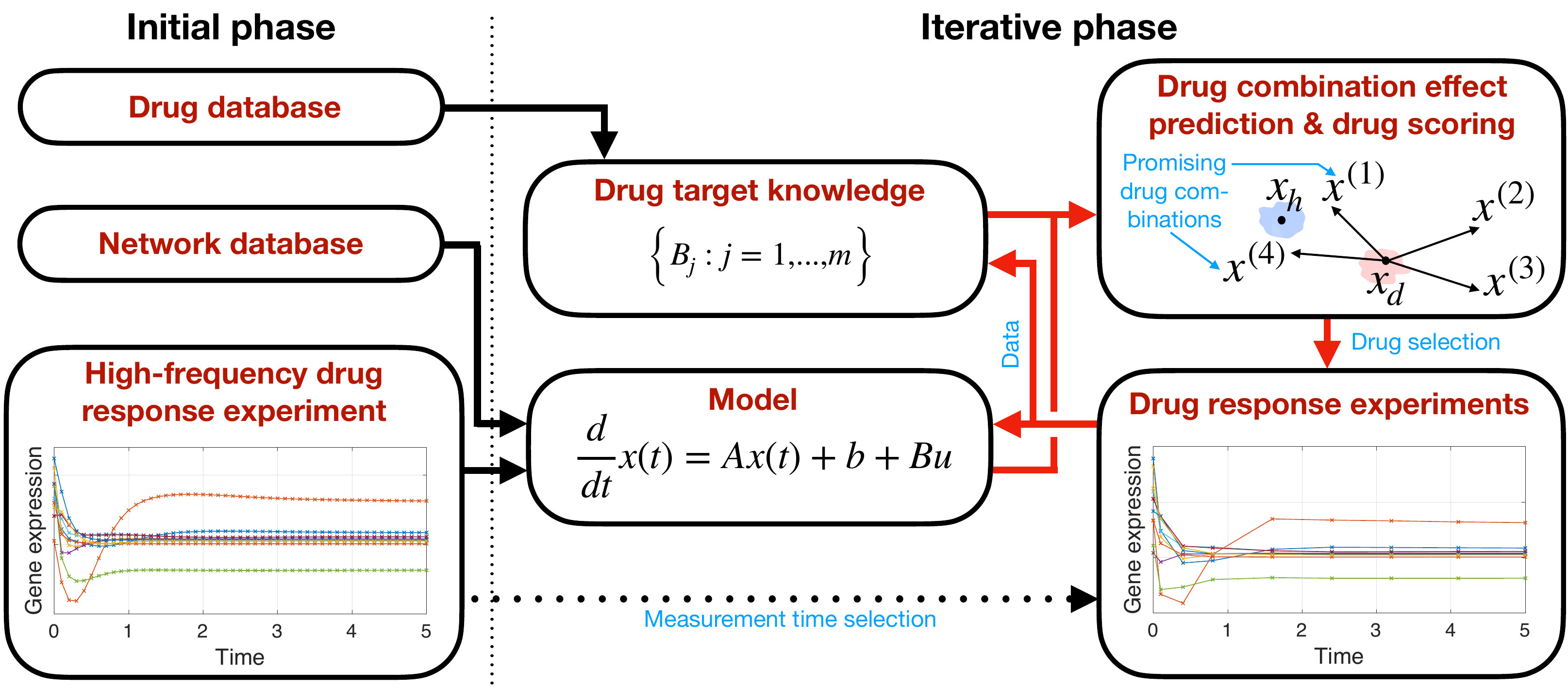}
      \caption{Pipeline of the model-based experimental cycle. The iterative modelling-experiment cycle is illustrated with red arrows.}
      \label{fig:pipeline}
   \end{figure*}

\section{PROBLEM FORMULATION}

We consider linear systems with dynamics given by
\begin{equation} \label{eq:model}
\frac{d}{dt}x(t) = Ax(t) + b + Bu, \quad x(0) = x_0
\end{equation}
where $x(t) \in \mathbb{R}^n$ is the system's gene expression state, $(A,b)$ determines the disease system, $B \in \mathbb{R}^{n \times m}$ is the drug effect matrix with $m$ the number of different drugs considered, and $u$ represents the dosage for each drug. In the drug experiments where a drug is applied to the organoid in a disease state, it is reasonable to assume that the system is initialised from the drug-free disease steady state, that is, $x_0 = x_d = -A^{-1}b$. The drugs are applied at time zero, and $u$ is assumed to be constant for $t > 0$. Moreover, the goal is to find optimal $K$-drug combinations for different values of $K$, that is, to consider inputs from the set $\mathcal{S}(K) := \{u \in \mathbb{R}_+^{m} \ : \ \|u\|_0 \le K \}$. As the input $u$ is related to drug dosage, negative values are excluded. In reality, there should also be an upper bound for dosage, but in this simplified study, we do not impose upper bounds, since the numerical values in the drug effect matrices are rather arbitrary.

The system $(A,b)$ is initially unknown, but the topology  of the underlying gene regulatory network (zero-structure of $A$) is assumed to be known. Based on knowledge of drug targets, it is assumed that the zero-structure of $B$ is (partially) known, as well as the signs of these known effects. In addition, the drugs' primary target genes are known, and these effects are assumed to be stronger than those on other genes. For details, see §\ref{sec:init} on how the estimated $B$-matrix is initialised.

The goal is to select a combination (and dosages) of drugs that steer the system's steady state towards a given healthy state $x_h$. The steady state corresponding to a drug combination encoded in an input vector $u$, is given by $-A^{-1}(b + Bu)$, and in principle, the problem could be formulated as a minimisation problem $\min_{u} \|-A^{-1}(b + Bu) - x_h\|$. However, using the inverse of a poorly known matrix $A$ in the cost function definition may lead to amplification of model uncertainties. Therefore, we instead formulate the problem using directly the right-hand side of the model \eqref{eq:model}, and try to find a drug combination that would make $x_h$ the steady state of the system, that is, $Ax_h + b + Bu = 0$. We formulate the associated minimisation problem as
\begin{equation} \label{eq:cost}
\min_{u \in \mathcal{S}(K)} J(u) \quad \textup{where} \quad J(u) = \|Ax_h + b + Bu\|_W^2
\end{equation}
where the weight matrix has the form $W = I + \gamma^2pp^{\top}$ and $p \in \mathbb{R}^n$, with $\|p\|=1$ defining a direction in the state space associated with the disease phenotype, that is, a direction in which deviations are particularly unwanted. Note that the term inside the cost function can be rewritten as $Ax_h+b+Bu = A(x_h-x_d) + Bu$.

For convenience, let us define the cost function on the index space, that is, for a $K$-drug combination $C$, denote
\begin{equation} \label{eq:indCost}
F(C) = \min_{u \in \mathcal{S}(K)} J(u) \quad \textup{s.t. } u_j=0 \ \textup{if } j \notin C.
\end{equation}
That is, $F(C)$ gives the optimal value for the cost function $J$ corresponding to the drug combination $C$.

The computational procedure is tailored for an approach utilising midbrain organoid cultures in the context of Parkinson's disease \cite{zagare}. Several organoids can be cultivated in parallel, and therefore in one experimental cycle, several different drugs can be tested. We assume that there is capacity to test five drugs at a time, and in total, five experimental cycles can be carried out. Testing a drug $i$ means collecting noisy time series data $y_j = x(t_j) + v_j$ where $x$ satisfies \eqref{eq:model} with $u_i=1$. The measurement (bulk RNA-sequencing) kills the sample, and hence a separate organoid is required for each time point in the time series data. Process noise is therefore excluded from the dynamics in \eqref{eq:model}, since the noise terms would anyway be different for different samples. The resulting data are used for improving the estimate of the model $(A,b)$ as well as the column $i$ in the matrix $B$. The challenge is to choose the drugs to test based on the knowledge of the model $(A,b)$ and the incomplete prior knowledge of the drug effects $B$.

\section{METHODS}

The pipeline for the iterative cycle is illustrated in Fig.~\ref{fig:pipeline}. On each iteration, the model is used together with the knowledge on drug effects to evaluate different drug combinations. This evaluation is used to select drugs for experimental testing. Data from the drug tests are used to improve the model and the drug effect knowledge.

In this section, $k$ denotes the iteration number. Variables $A^{(k)}$, $b^{(k)}$, and $B^{(k)}$ denote estimated parameters during the $k^{\textup{th}}$ iteration (based on data collected in the first $k-1$ iterations). It is infeasible to even computationally tract all possible drug combinations. Therefore we define the set $\mathcal{C}^{(k)}$ of drug combinations to keep track of. The set is extended as the iterations proceed. We also denote by $F^{(k)}$ the cost function defined in \eqref{eq:indCost}, but where the current estimates $A^{(k)}$, $b^{(k)}$, and $B^{(k)}$ are used instead of $A$, $b$, and $B$.

\subsection{Initialisation}
\label{sec:init}

The initial phase of the procedure involves a high-frequency experiment with densely sampled time points on one preferred drug known to elicit a strong response. The data from this initial experiment are used to estimate an initial model $A^{(1)}$ as explained in §\ref{sec:id}. In the identification, it is not possible to distinguish $b$ from the drug effect $B_i$. Assuming that the initial state is the drug-free steady state $x_d$, and that it is also measured, that is, $y_0 = x_d+v_0$, then $b$ can be estimated as $b = -A^{(1)}y_0$. Multiple experiments can be averaged to improve this estimate. 

The prior estimate of $B$ (before the high frequency experiment), denoted by $B^{(0)}$, originates from a drug database. The drug database contains drug target genes and the direction of the effects (activation/inhibition). In addition, we assume that the primary target genes are known. Each column in $B$ corresponds to a drug in the library. The entry in the column $B_i^{(0)}$ corresponding to the primary target gene is set to 1 or -1. Entries corresponding to other target genes are set to 1/2 or -1/2.

The high-frequency experiment studies the response to one drug $i$. The data from this experiment are then used to estimate the corresponding column $B_i$. This results in the matrix estimate $B^{(1)}$ where other columns are as in $B^{(0)}$.

The set of interesting drug combinations $\mathcal{C}^{(1)}$ is initialised by including all individual drugs and all two-drug combinations. For each drug combination $C \in \mathcal{C}^{(1)}$, the cost $F^{(1)}(C)$ is evaluated.

\subsection{Adding drug combinations to $\mathcal{C}^{(k-1)}$}
\label{sec:Sk}

Starting from the second iteration, the set of interesting drug combination $\mathcal{C}^{(k-1)}$ is extended by adding new combinations. The top 1000 combinations in $\mathcal{C}^{(k-1)}$ are selected and one drug $i$ at a time is added to each of the top combinations $C_j \in \mathcal{C}^{(k-1)}$, and $F^{(k)}(C_j \cup \{i\})$ is calculated. Top 1000 new combinations found are then included to form $\mathcal{C}^{(k)}$. With this procedure, $\mathcal{C}^{(k)}$ contains drug combinations with at most $k+1$ drugs.

\subsection{Drug scoring}

It is not directly obvious which drugs should be selected for experimental testing. In theory, the lowest cost function value is obtained with the highest number of drugs in a combination. However, if one were to select all drugs from the best found five-drug combination, for example, it would eat up the entire testing capacity of one iteration, and it is likely that not all of these five drugs are equally important. 

To select the drugs for experimental testing, we have devised a two-stage process. First, a set of candidate drugs for testing is formed by including the top five single drugs that have not been tested in earlier iterations. Then we add drugs from the top two-drug combinations until (at least) five non-tested drugs have been added. Then top three-drug combinations are scanned, etc. Second, to choose the drugs for testing from this candidate set, a drug scoring scheme is used that is based on their importance in different drug combinations. The importance score for drug $c_j$ in a $K$-drug combination $C = \{c_1,...,c_K\}$ is defined as the relative improvement in the cost function value when drug $c_j$ is included in the combination:
\[
1-\frac{F^{(k)}(C)}{F^{(k)}(C\setminus \{c_j\})}.
\]
The final importance score for a drug is defined as the maximum of the importance scores over all combinations in $\mathcal{C}^{(k)}$ where the drug is included.

\subsection{Continuous-time identification}
\label{sec:id}

The experimental testing of a drug produces time series data $y_j = x(t_j) + v_j$ where $v_j$ is measurement noise. As discussed, each time point in the collected time series data requires an individual organoid to be grown and sacrificed for the measurement. Each time point therefore comes with a significant cost, and as a result, the time series data are very sparsely sampled. Moreover, the sampling times are not equally spaced, in order to optimally capture the transient drug response. Data from the initial high-frequency experiment are used to determine the temporal dynamics of the drug response and to select good sampling times for the other experiments. In this work, however, this is done manually rather than by maximising some information content functional.

Due to the sparse and irregular sampling, and to accurately preserve the prior knowledge on the underlying regulatory topology, it is better to directly identify a continuous-time model from the sampled data \cite{contIDbook}. Integrating \eqref{eq:model} gives
\[
x(t+\Delta t) = x(t) + \int_0^{\Delta t} Ax(t+s) ds + \Delta t b
\]
and, further, 
\begin{equation}
    \label{eq:exp}
    \frac{x(t+\Delta t) - x(t)}{\Delta t} = A \frac{1}{\Delta t} \int_0^{\Delta t} x(t+s) ds + b.
\end{equation}
Note that this is not an approximation. Approximating $x(t+s) \approx x(t)$ in the integrand would lead to the explicit Euler scheme. However, we use an iterative scheme for estimating the state trajectory and the system $(A,b)$. 

The iteration is initialised by $A_0=0$, $b_0 = 0$ (here we omit the main iteration superscript index and use the subscript to indicate the iteration inside the continuous-time identification scheme). On the iteration $i$, the model $(A_{i-1},b_{i-1})$ is used to approximate $x(t)$ piecewise on intervals $[t_j,t_{j+1})$ by propagating $y_j$ through the model \eqref{eq:model}. This piecewise defined trajectory is then integrated over the sampling intervals to calculate the right-hand side of \eqref{eq:exp} to form the regression problem with $\frac{y_{j+1}-y_j}{t_{j+1}-t_j}$ on the left-hand side. Data from all experiments are then concatenated, and $(A_i,b_i)$ are solved from the regression problem by enforcing the known topology of $A$. It should be noted that on the first iteration when $A_0=0$, $b_0 = 0$, $x(t)$ is piecewise constant, and the approximation corresponds precisely to the explicit Euler discretisation.

This iteration can be interpreted as a fixed-point iteration. It converges relatively quickly. We run ten identification iterations on each main iteration cycle $k$. The estimated $A^{(k+1)}$, $b^{(k+1)}$, and $B^{(k+1)}$ are obtained by concatenating all data from the initial high-frequency experiment and the drug experiments conducted so far.

\subsection{Final search}

The full experiment iteration is done multiple times, depending on available resources. In the experimental results of this article §\ref{sec:result}, five iterations are run, with five drugs tested on each iteration.

After the iterative phase is completed, a final search for the optimal $K$-drug combinations is performed. This final search goes through all $K$-drug combinations that can be formed from the 26 drugs that have been experimentally tested (1 in the initial experiment and 5 per iteration in the 5 iterations). The cost $F^{(6)}(C)$ is evaluated for all combinations $C$ and the best one is chosen for evaluation.

\subsection{Data generation}

The ground truth system $(A,b,B)$ and the prior drug knowledge $B^{(0)}$ are randomly generated as explained in this section. We generate 20 random replicates, and run the procedure on each replicate.

\begin{figure}
    \centering
    \includegraphics[width=\linewidth]{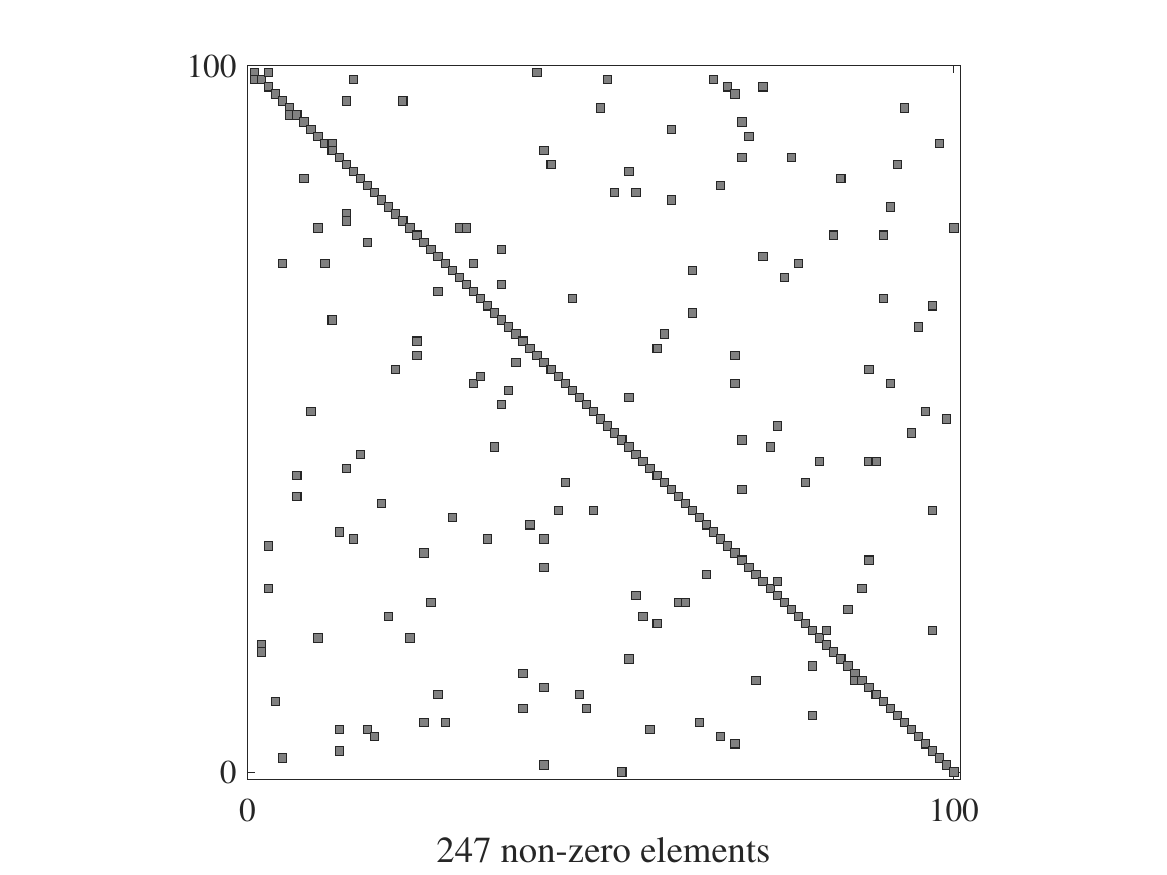}
    \caption{A typical topology generated during one replicate.}
    \label{fig:spyA}
\end{figure}

The regulatory topology $G$ of the $100 \times 100$ $A$-matrix is randomly constructed\footnote{$G$ is the underlying gene regulatory network topology, that is, the off-diagonal nonzero-structure of $A$.}. 
First, 50 links are formed between two randomly drawn genes. Then, going through genes one at a time, $i=1,...,100$, if $\sum_{j=1}^{100}G_{i,j}=0$, one gene $j \ne i$ is chosen randomly, and a link is added by setting $G_{i,j}=1$. Then, if $\sum_{j=1}^{100}G_{j,i}=0$, similarly we set $G_{j,i}=1$ for one random gene $j \ne i$. The resulting network is fully connected and every gene is regulated by at least one other gene and each gene regulates at least one other gene. A typical topology thus formed is shown in Fig.~\ref{fig:spyA}.  

When the topology has been generated, the non-zero non-diagonal entries in the $A$-matrix are drawn from $U\big((-10,-1)\cup (1,10)\big)$. The diagonal entries are drawn from $U(-20,-2)$. If the matrix is unstable ($A$ has an eigenvalue with positive real part), the diagonal entries are re-drawn.

The entries of the vector $b$ are drawn from $U(0,10)$ and the vector $p$ is drawn from a uniform distribution on the unit ball and $\gamma = 10$.

The drug effect matrix $B$ is generated by randomly choosing 6\% of all entries of the $100\times 200$ matrix, and drawing the values of these entries from $U\big((-10,-1)\cup (1,10)\big)$. To then construct the ``drug database'', for each drug (column), the entry with maximal absolute value is determined, and the corresponding entry in $B^{(0)}$ is set to 1 or -1 depending on the sign of the entry in $B$. Then, we go through the remaining non-zero values on the column $B_j$, and with 50\% probability, the corresponding entries in $B^{(0)}$ are set to 1/2 or -1/2.

The healthy target state is defined as $x_h = -DA^{-1}(b + \bar B \bar u)$ where $\bar B$ contains only the primary effects of drugs (on each column the entry with largest absolute value), and $\bar u_j \sim U(0.5,1.5)$ for $j=1,2,3,4$, and $\bar u_j = 0$ for $j>4$. $D$ is a diagonal matrix whose diagonal entries are drawn from $\mathcal{N}(1,0.01^2)$. The idea of this definition is to have a kind of a ground truth drug combination that is used to generate the data, but the secondary effects of the true drugs are considered as (unwanted) side effects. However, since the secondary effects are omitted in the definition, the drug combination 1,2,3,4 does not typically emerge as the optimal combination. Drug 1 is used in the high-frequency experiment since this drug should have an effect in the desired direction, bu the construction of $x_h$. In reality, the drug for the initial experiment is chosen using some prior knowledge on the disease and the drugs.

The data are simulated using the explicit Euler method with $\Delta t = 0.01$. Each trajectory is initialised from the drug-free steady state $x_d = -A^{-1}b$. Each time point is corrupted by noise $v_j\sim \mathcal{N}(0,0.005^2I)$. 
 The high-frequency dataset consists of 51 time points with equal spacing 0, 0.1,..., 5.0. All low-frequency experiments share the time sequence with the sampling becoming less dense at later times, in order to better capture the initial transient response. The nine measurement times are 0, 0.1, 0.4, 0.8, 1.6, 2.4, 3.2, 4.1, 5.0. The time series shown in Fig. \ref{fig:pipeline} are generated by the described procedure (10 genes shown).

\section{RESULTS}

\subsection{Model estimation}

\begin{figure}
    \centering
    \includegraphics[width=\linewidth]{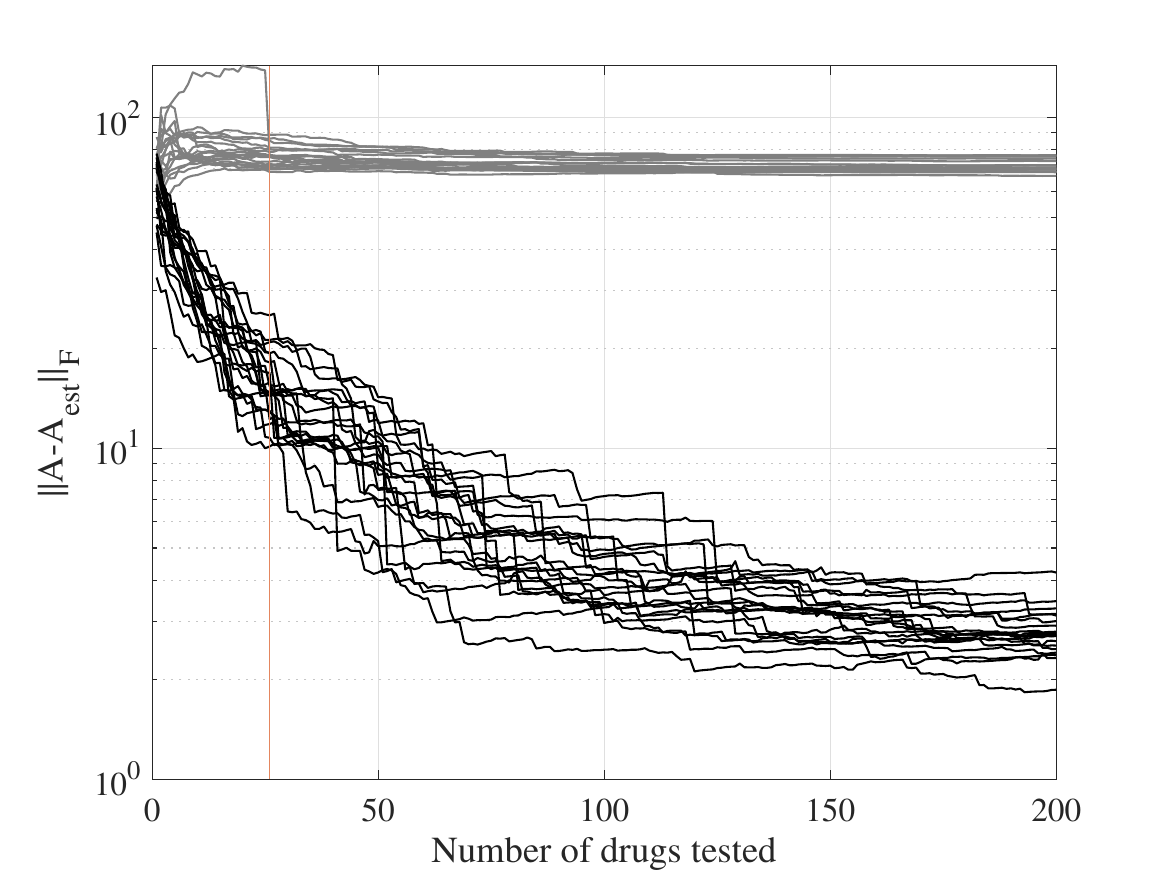}
    \caption{Performance of the estimation methods across 20 replicates measured by the Frobenius norm of the error between the estimated $A$ and true $A$.
    Black: iterative continuous-time scheme. Gray: explicit Euler discretisation.
    The vertical red line indicates the  drug testing capacity used in the simulation (1 high and 25 low frequency tests) in §\ref{sec:result}.}
    \label{fig:perfexp}
\end{figure}

Figure~\ref{fig:perfexp} demonstrates the performance of the continuous-time estimation method compared to the explicit Euler scheme across 20 replicates.
For demonstration purposes, we pick all the 200 drugs in the drug bank sequentially to excite the system.
Since the simulations are initialised from the drug-free steady state $x_d$, and since each drug affects only a handful of genes, the information content in each individual time series is not very high.

The explicit Euler discretisation performs poorly, as the gray curves invariably rise after the initial high-frequency time series sequence change to the low-frequency ones, which are inadequate for approximating the time derivatives and cause a large asymptotic bias.
On the contrary, the errors of the iterative continuous-time method represented by the black curves show an overall downward trend as more time-series data come in after each new drug test, and the asymptotic error is an order of magnitude smaller than that of the explicit Euler scheme.
We note that the error appears to converge to a nonzero value possibly due to the correlation bias induced by introducing the same noise in both the regressor and the outcome side of the regression model, even though the measurement noise of different time point is assumed to be i.i.d.
In principle, one way to remove this bias is to use the instrumental variable approach \cite{riva,contIDbook}, however the success hinges on finding a suitable independent variable to decorrelate the noise on both sides of~\eqref{eq:exp}.
In practice, we find that emulating the trajectory $x(t)$ on $[t_j,t_{j+1})$ using $y_{j-1}$ instead of $y_j$ as the starting point (as the noise $v_{j-1}$ is independent from $v_j$) yields no discernible improvement.

\subsection{Drug combination discovery}
\label{sec:result}

\begin{figure*}
  \centering
  \includegraphics[width=.49\linewidth]{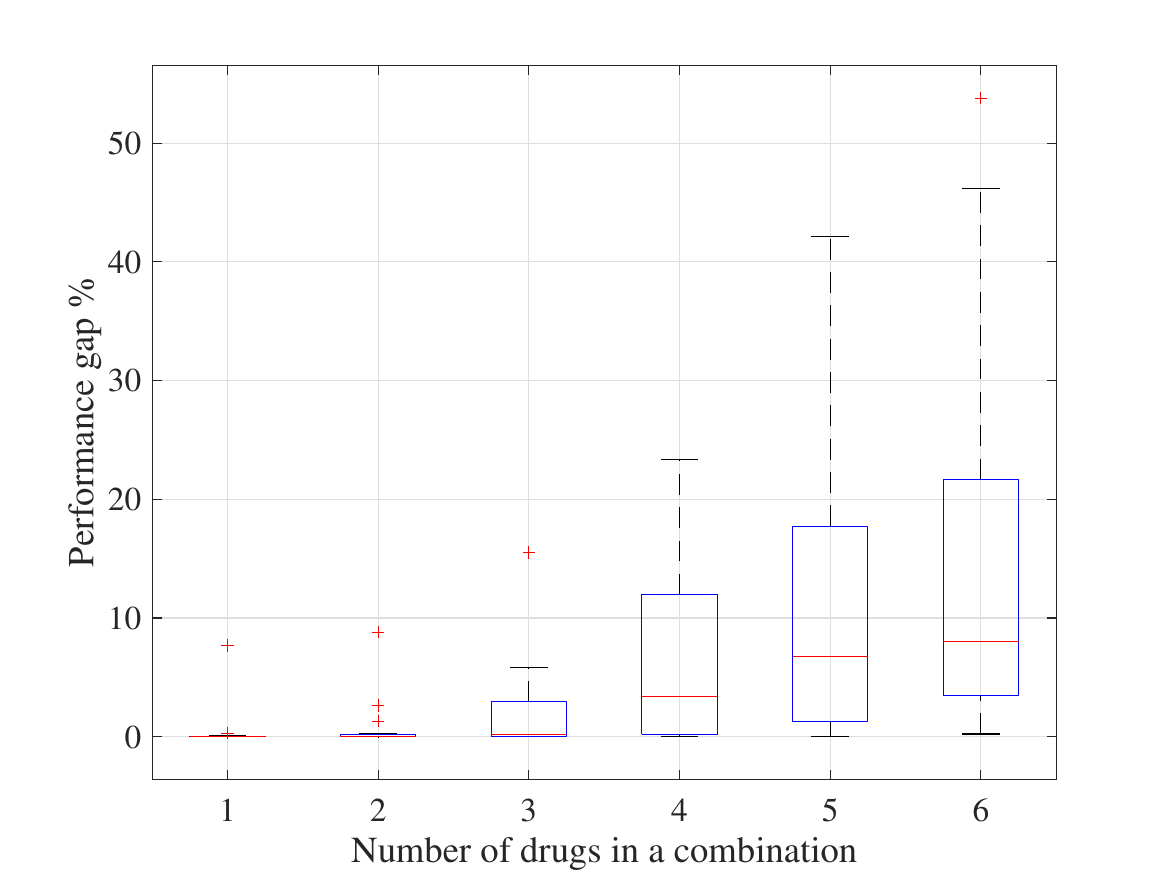}
  \includegraphics[width=.49\linewidth]{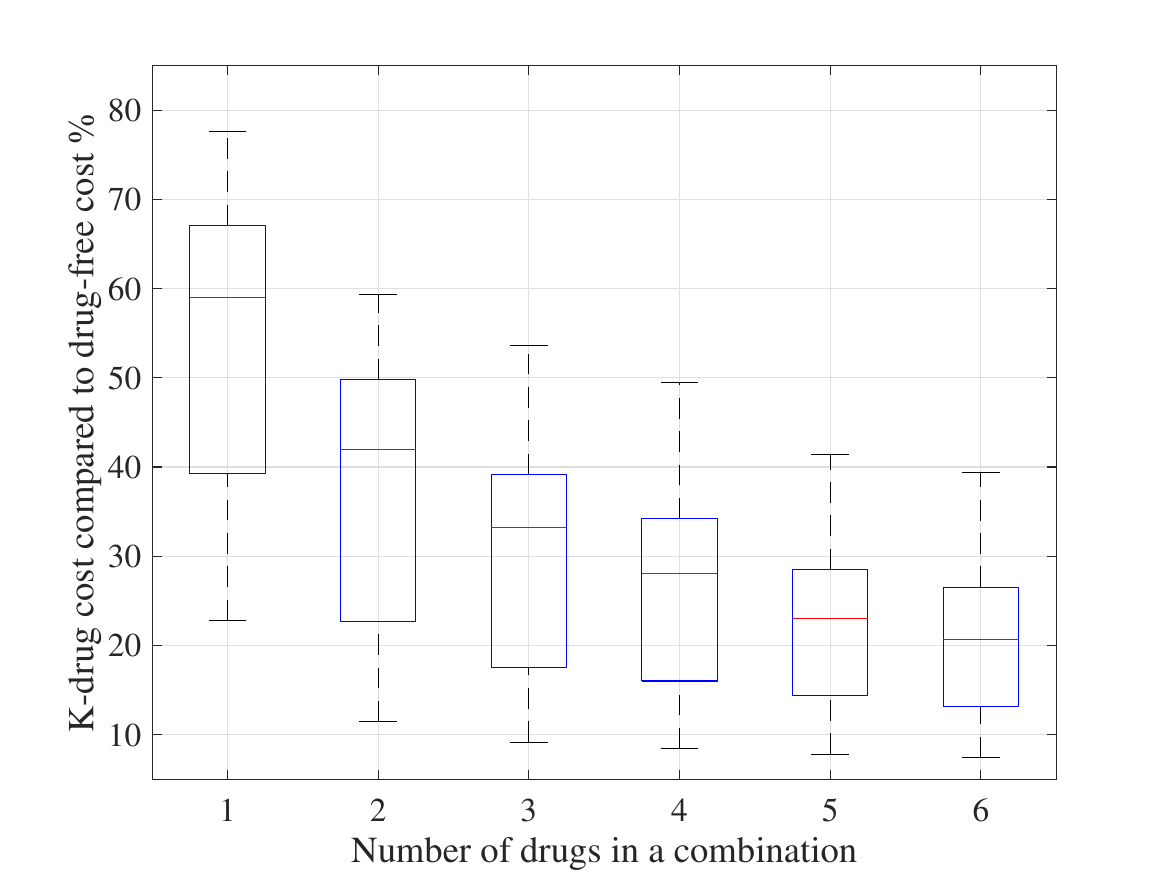}
  \caption{Left: performance gap between the best found drug combinations and the true optima.
  Right: the costs for the best found combinations compared to the disease state cost.
  Both measures are computed for 1 to 6-drug combinations and the statistics across 20 replicates are presented in box plot format.}
  \label{fig:performance31}
\end{figure*}

To illustrate how the proposed computational procedure discover and select prospective drug repositioning candidates, we show two key performance indicators across 20 replicates in Fig.~\ref{fig:performance31}.
The statistics gathered across the 20 replicates are shown in the box plot format. 

The left panel shows the performance gaps between the best found $K$-drug combination and dosage $u_K$, and the ``true'' optimum $\hat J_K := \min_{u \in \hat{\mathcal{S}}(K)} J(u)$, for $K=1,...,6$. The performance gap is defined as $\frac{J(u_K)-\hat J_K}{\hat J_K}$. Note that finding the true optimum $\hat J_K$ is a combinatorial problem, and we solve it in the iteratively constructed search space $\hat{\mathcal{S}}(K)$ using the same search procedure as in the introduced procedure (§III-A-B), but using always the ground truth cost $F(\cdot)$ to evaluate drug combinations. Moreover, we include 3000 $K$-drug combinations ($K>2$) to be considered for $\hat J_K$, instead of 1000 used otherwise in the procedure (see §III-B).
Note also that the estimated best drug combination may differ from that found using the true model.

The performance gap is nearly zero in most replicates for $K \le 2$. The near-zero median performance gaps indicate the efficacy of the iterative continuous-time estimation method, required for a good estimation of the drug dosages, and the precision in finding the most essential drugs for testing.
When $K > 2$, the combinatorial nature of the problem contrasted with the limited testing resources becomes more significant, leading to an increase in the median performance gap. However, it still remains below 10\%, which can be considered satisfactory.

The right panel shows the ratio between the cost of the best found drug combination and the cost of taking no drugs at all: $\frac{J(u_K)}{J(0)}$. The smaller the values, the better are the potential health benefits.
There is a nearly 20\%pt. median improvement by taking the recommended 2-drug combination rather than only 1 drug. As $K$ increases, the estimated costs decrease as expected, though the improvement becomes less marked.

\section{CONCLUSIONS AND DISCUSSION}

We presented an iterative procedure for drug repositioning studies that aims for quantitative estimation of drug responses, that, in turn, enables analyses on drug combinations. In a pilot study using simulated data, the procedure was able to identify drug combinations of up to six drugs with a performance gap below 10\% compared to true optima. This work illustrated the feasibility of using dynamical models coupled with system identification and iterative feedback to accelerate drug discovery. In future work, we plan to apply this procedure to real world problems, in particular, to repurposing drugs in Parkinson’s disease models in midbrain organoids.

The presented study has its limitations. Firstly, gene expression dynamics are certainly not linear, neither is the drug response. Indeed, the effect of a drug typically plateaus as the dosage is increased, and often the nominal dosage is already chosen with the drug's desired effect and potential unwanted side effects in mind. Moreover, drugs can also affect each other when taken in combination, and certain combinations can even be harmful for the organism. Some of these harmful combinations are already known and are registered in the drug databases. It is possible to move to nonlinear modeling, for example using the approach of our earlier work \cite{bingo}. However, while nonlinear models may benefit from increased realism and accuracy, their fitting requires more data, so improved performance is not necessarily guaranteed.

Another limitation is the assumption on known topology of the underlying gene regulatory network. This can be somewhat circumvented by relaxing the topology enforcement by an appropriate regularisation mechanism with the regularisation penalty modulated according to prior knowledge. Obtaining real data will help assess the severity of these limitations and to guide further development.

In the procedure, the drugs are selected for experimentation solely based on their estimated utility in steering the system. Another way would be to balance between exploration and exploitation by trying to maximise the information obtained from the resulting data, that is, to select drugs targeting genes whose downstream dynamics are not yet well known.

Due to the experimental setup, there is also a tradeoff between the number of drugs tested and the number of time points in each time series. In theory, the drug effect alone could be inferred from a steady-state measurement $y_{\textup{ss}}$ of the drug response by $B_i = A(x_d-y_{\textup{ss}})$ if $u_i=1$. For some drugs, therefore, it might not be necessary to collect that many intermediary time points, and instead use the saved resources to increase the number of tested drugs.

In the construction of the prior drug effect knowledge, we here considered knowledge only on lists of target genes. However, there exist databases on quantitative transcriptional drug effects \cite{L1000} instead of the qualitative data we use in this study. The prior knowledge could potentially be improved by integrating such knowledge, thereby saving resources. This database is used in \cite{grand} to form drug signatures on a regulatory network to be overlaid with disease signatures.

\end{document}